\documentclass[12pt,a4paper]{article}

\usepackage{epsfig}
\usepackage{amsmath,amsfonts,amssymb}
\usepackage{t1enc}
\usepackage{cite}

\providecommand{\openone}{\leavevmode\hbox{\small1\kern-3.8pt\normalsize1}}

\newcommand{\IM}{\text{Im}\,}

\newcommand{\smn}{\sigma^{\mu \nu}}

\newcommand{\gm}{\gamma^\mu}

\newcommand{\Gmna}{G_{\mu \nu}^a}
\newcommand{\la}{\lambda^a}

\newcommand{\gul}{\zeta_{ut}^L}
\newcommand{\gur}{\zeta_{ut}^R}
\newcommand{\gcl}{\zeta_{ct}^L}
\newcommand{\gcr}{\zeta_{ct}^R}
\newcommand{\gql}{\zeta_{qt}^L}
\newcommand{\gqr}{\zeta_{qt}^R}

\newcommand{\atu}{\alpha_{tu}}
\newcommand{\aut}{\alpha_{ut}}
\newcommand{\btu}{\beta_{tu}}
\newcommand{\but}{\beta_{ut}}
\newcommand{\atc}{\alpha_{tc}}
\newcommand{\act}{\alpha_{ct}}
\newcommand{\btc}{\beta_{tc}}
\newcommand{\bct}{\beta_{ct}}
\newcommand{\atq}{\alpha_{tq}}
\newcommand{\aqt}{\alpha_{qt}}
\newcommand{\btq}{\beta_{tq}}
\newcommand{\bqt}{\beta_{qt}}

\begin{document}

\begin{center}
\begin{Large}
{\bf $Zt$, $\gamma t$ and $t$ production at hadron colliders \\[1mm]
via strong flavour-changing neutral couplings}
\end{Large}

\vspace{0.5cm}
J. A. Aguilar--Saavedra  \\[0.2cm] 
{\it Departamento de Física Teórica y del Cosmos and CAFPE, \\
Universidad de Granada, E-18071 Granada, Spain} \\[0.1cm]
\end{center}

\begin{abstract}
We consider $gq \to Zt$, $gq \to \gamma t$ and $gq \to t$ production ($q=u,c$) mediated by strong flavour-changing neutral interactions within an effective operator framework.
We provide total cross sections for Tevatron and LHC, showing explicitly that the six processes can be described in full generality in terms of only two parameters (anomalous couplings) for $q=u$ plus two for $q=c$.
In our work we take into account and study in detail the effects of top quark decay.
For $\gamma t$, the inclusion of the top quark decay in the matrix element reveals an striking result: the largest contribution to the final state, {\em e.g.} $\gamma \ell \nu b$ with $\ell = e,\mu,\tau$, does not result from $gq \to \gamma t \to \gamma \ell \nu b$ but from on-shell $g q \to t$ production with $t \to \gamma \ell \nu b$, being the photon radiated off the top decay products. This contribution, missed in previous literature, increases the signal cross sections by factors ranging between 3 and 6.5.
\end{abstract}

\section{Introduction}

The study of the top quark properties provides an excellent opportunity to probe new physics beyond the Standard Model (SM)~\cite{Beneke:2000hk,Bernreuther:2008ju}, in particular for those models in which the top quark plays a special role~\cite{Hill:2002ap}, is composite~\cite{Giudice:2007fh,Agashe:2009di} or has a sizeable mixing with a heavy partner $T$~\cite{delAguila:1998tp,delAguila:2000aa,Agashe:2006wa}. If new physics exists at a scale above few TeV, the new states may be too heavy to be produced even at the Large Hadron Collider (LHC), and the only observable effects may be the indirect ones, {\em i.e.} the modification of SM particle properties. In this respect, top flavour-changing neutral (FCN) interactions provide a unique window to new physics, because they are extremely suppressed within the SM~\cite{Eilam:1990zc,Mele:1998ag,AguilarSaavedra:2002ns}.

For the study of indirect effects of new physics above the electroweak symmetry breaking scale, the framework of gauge-invariant effective operators~\cite{Burges:1983zg,Leung:1984ni,Buchmuller:1985jz} is the most convenient one. Within this philosophy, new physics is described by an effective Lagrangian expanded in inverse powers of the new physics scale $\Lambda$,
\begin{equation}
\mathcal{L}^\text{eff} = \sum \frac{C_x}{\Lambda^2} O_x + \dots \,,
\label{ec:effL}
\end{equation}
where $O_x$ are dimension-six gauge-invariant operators and $C_x$ are complex constants. The dots in the above equation stand for higher-dimension operators neglected in this work, whose contributions are suppressed by higher powers of $\Lambda$. Any new physics contribution (up to higher-order terms) can be written in terms of an operator basis $\{O_x\}$ such as the one given in Ref.~\cite{Buchmuller:1985jz}. Here we will focus on top strong FCN interactions with the gluon. There are three operators listed in Ref.~\cite{Buchmuller:1985jz} mediating such interactions,
\begin{align}
& O_{uG\phi}^{ij} = (\bar q_{Li} \la \smn u_{Rj}) \tilde \phi \, \Gmna \,, \notag \\
& O_{qG}^{ij} = \bar q_{Li} \la  \gm D^\nu q_{Lj} \Gmna \,, \notag \\
& O_{uG}^{ij} = \bar u_{Ri} \la \gm D^\nu u_{Rj} \Gmna \,,
\label{ec:Oall}
\end{align}
with $i,j=1,2,3$ flavour indices (see the next section for notation). However,
the latter two were found to be redundant in Refs.~\cite{AguilarSaavedra:2008zc}. Therefore, top FCN interactions with the gluon can be described by a single operator $O_{uG\phi}^{ij}$, with flavour indices $i,j=1,3/3,1$ for $gtu$ and $i,j=2,3/3,2$ for $gtc$. This simplification provides an enormous advantage for phenomenology, reducing the number of effective operator coefficients necessary to parameterise observables such as FCN cross sections and branching ratios to only two for the up quark and two for the charm.

In this paper we study $t$, $\gamma t$ and $Zt$ production at hadron
colliders~\cite{Hosch:1997gz,Tait:1997fe,delAguila:1999ac,delAguila:1999ec,
Cao:2007dk,Ferreira:2005dr,Ferreira:2006in}, mediated by the first operator in Eqs.~(\ref{ec:Oall}). These processes have no SM irreducible backgrounds and their observation would then be a clear signal of top flavour violation.\footnote{Top FCN couplings can also mediate $tj$ production, {\em e.g.} $gq \to gt$, $qq \to qt$~\cite{Ferreira:2006xe,Han:1998tp,Gao:2009rf}, with a total cross section about half the one for $gq \to t$~\cite{Gao:2009rf}. These processes, however, give a signal consisting of a top quark plus a (forward) jet, for which $t$-channel single top production constitutes an irreducible background.}
The production of $t$, $\gamma t$ and $Zt$ through strong FCN couplings has been previously considered~\cite{Ferreira:2005dr,Ferreira:2006in} but, as we will show, important simplifications and improvements are possible from the theoretical point of view.
We will calculate their production cross sections at LHC and Tevatron. It will be explicitly shown that previous results obtained using a non-minimal operator basis~\cite{Ferreira:2005dr,Ferreira:2006in,Ferreira:2006xe} which involves more parameters can be easily reproduced with a simple change of variables.
We will also clarify the relations among the total cross sections for these processes and the top decay width for $t \to qg$, which result from symmetries. Moreover,
in our study we take into account the top quark decay, implementing the processes in the generator {\tt Protos}~\cite{AguilarSaavedra:2008gt}. For $g q \to \gamma t$ this proves to be essential. When the top quark decay is consistently included in the matrix element, it is found that the largest contribution to the cross section actually results from the diagrams corresponding to on-shell $gq \to t$ production with the photon radiated off the top quark decay products. This can be easily understood, given the large cross section for direct $t$ production~\cite{Hosch:1997gz,Ferreira:2005dr} and the sizeable branching ratio for $t \to \gamma Wb$~\cite{Mahlon:1998fr,Mele:1999zx}. The increase in the cross sections, up to a factor of six, depends on the centre of mass (CM) energy and the initial state, and significantly enhances the experimental sensitivity to $gtu$, $gtc$ couplings in this process.

The rest of this paper is organised as follows. In section~\ref{sec:2} we set our notation and work out in detail the relation between our minimal operator basis and that of Refs.~\cite{Ferreira:2005dr,Ferreira:2006xe,Ferreira:2006in}. In section~\ref{sec:3} we study top FCN decays $t \to qg$ and direct $t$ production. Sections~\ref{sec:4} and \ref{sec:5} are devoted to $\gamma t$ and $Zt$ production, respectively. We summarise our results in section~\ref{sec:6}.

\section{Parameterisation of strong top FCN interactions}
\label{sec:2}

As we have pointed out in the introduction, the only independent dimension-six operators contributing to the $gtu$, $gtc$ vertices are
\begin{align}
& O_{uG\phi}^{ij} = (\bar q_{Li} \la \smn u_{Rj}) \tilde \phi \, \Gmna \,,
\label{ec:Ostr}
\end{align}
with flavour indices $i,j=1,3/3,1$ for $gtu$ and $i,j=2,3/3,2$ for $gtc$. 
Here $q_{Li}$ and $u_{Rj}$ are the quark weak interaction eigenstates, $G_{\mu \nu}^a$ the gluon field tensor, $\phi$ the SM Higgs doublet with a vacuum expectation value $v=246$ GeV, $\tilde \phi = i \tau^2 \phi^*$, $\tau^I$ the Pauli matrices and $\la$ the Gell-Mann matrices (see Refs.~\cite{AguilarSaavedra:2008zc} for further notation).
These operators generate the vertex
\begin{equation}
\mathcal{L}_{gtu} = - g_s \bar u \, \la \frac{i \smn q_\nu}{m_t} \left( \gul P_L + \gur P_R \right) t\; G_\mu^a + \mathrm{h.c.} \,,
\label{ec:gtu}
\end{equation}
and a similar one for $gtc$, involving anomalous couplings $\gcl$ and $\gcr$. The relation between anomalous couplings and effective operator coefficients is
\begin{align}
& \gul = \frac{\sqrt 2}{g_s} C_{uG\phi}^{31*} \frac{v m_t}{\Lambda^2} \,,
&& \gur = \frac{\sqrt 2}{g_s} C_{uG\phi}^{13} \frac{v m_t}{\Lambda^2} \,, \notag \\
& \gcl = \frac{\sqrt 2}{g_s} C_{uG\phi}^{32*} \frac{v m_t}{\Lambda^2} \,,
&& \gcr = \frac{\sqrt 2}{g_s} C_{uG\phi}^{23} \frac{v m_t}{\Lambda^2} \,.
\label{ec:ztoC}
\end{align}
On the other hand, the Lagrangian used in
Refs.~\cite{Ferreira:2005dr,Ferreira:2006in} to parameterise these interactions includes more effective operators,
\begin{equation}
\mathcal{L}'_{gtu} =
i \frac{\atu}{\Lambda^2} O_{uG}^{31} + i \frac{\aut}{\Lambda^2} O_{uG}^{13} + 
\frac{\btu}{\Lambda^2} O_{uG\phi}^{31} + \frac{\but}{\Lambda^2} O_{uG\phi}^{13} + \mathrm{h.c.}
\,,
\label{ec:L2}
\end{equation}
and a similar one $\mathcal{L}'_{gtc}$ for $gtc$ interactions, obtained by replacing $1 \to 2$ in the operator indices and $u \to c$ in the coefficient labels.
We can use the operator equalities~\cite{AguilarSaavedra:2008zc}
\begin{eqnarray}
O_{uG}^{3i} & = & \frac{1}{2 \sqrt 2 v} \left[ m_t O_{uG\phi}^{3i} - m_{u_i} ( O_{uG\phi}^{i3} )^\dagger \right] 
- \frac{8g_s}{9} O_{qu}^{(1,ki3k)} + \frac{g_s}{6} O_{qu}^{(8,ki3k)} \notag \\
& &  + \frac{g_s}{2} O_{uu}^{(8,3ikk)} + \frac{g_s}{4} O_{ud}^{(8,3ikk)}
\,, \notag \\
O_{uG}^{i3} & = & \frac{1}{2 \sqrt 2 v} \left[ m_{u_i} O_{uG\phi}^{i3} - m_t ( O_{uG\phi}^{3i} )^\dagger \right]
- \frac{8g_s}{9} O_{qu}^{(1,k3ik)} + \frac{g_s}{6} O_{qu}^{(8,k3ik)} \notag \\
& &  + \frac{g_s}{2} O_{uu}^{(8,i3kk)} + \frac{g_s}{4} O_{ud}^{(8,i3kk)} \,,
\label{ec:Oprel}
\end{eqnarray}
with $i=1,2$ (a sum over $k=1,2,3$ is understood) to write the operators $O_{uG}^{3i}$, $O_{uG}^{i3}$ in terms of other operators in the basis of Ref.~\cite{Buchmuller:1985jz}. The four-fermion operators entering these equalities are
\begin{align}
& O_{uu}^{(8,ijkl)} = \frac{1}{2} (\bar u_{Ri} \gm \la u_{Rj}) (\bar u_{Rk} \gamma_\mu \la u_{Rl}) \,, \notag \\
& O_{ud}^{(8,ijkl)} = (\bar u_{Ri} \gm \la u_{Rj}) (\bar d_{Rk} \gamma_\mu \la d_{Rl}) \,, \notag \\
& O_{qu}^{(1,ijkl)} = (\bar q_{Li} u_{Rj}) (\bar u_{Rk} q_{Ll}) \,, \notag \\
& O_{qu}^{(8,ijkl)} = (\bar q_{Li} \la u_{Rj}) (\bar u_{Rk} \la q_{Ll}) \,.
\end{align}
These operator relations result from the use of the gauge-invariant equations of motion, which amount to a field redefinition which leaves the path integral invariant
up to higher-order corrections~\cite{Georgi:1991ch,Arzt:1993gz}. Thus, the operator relations in Eqs.~(\ref{ec:Oprel}) are valid for any process, including those in which the top, light quark and/or gluon entering the FCN vertex are off-shell.
For the $gtu$, $gtc$ vertices the operator equalities in Eqs.~(\ref{ec:Oprel}) translate into relations between the operator coefficients in $\mathcal{L}'_{gtu}$, $\mathcal{L}'_{gtc}$ and those corresponding to the minimal set,
\begin{align}
& C_{uG\phi}^{31} = \btu + \frac{i m_t}{2 \sqrt 2 v} (\atu + \aut^*) \,, 
&& C_{uG\phi}^{13} = \but + \frac{i m_u}{2 \sqrt 2 v} (\aut + \atu^*) \,, \notag \\
& C_{uG\phi}^{32} = \btc + \frac{i m_t}{2 \sqrt 2 v} (\atc + \act^*) \,, 
&& C_{uG\phi}^{23} = \bct + \frac{i m_c}{2 \sqrt 2 v} (\act + \atc^*) \,.
\label{ec:trans}
\end{align}
Physical observables are the same under this change of variables plus some redefinitions of four-fermion operator coefficients. Nevertheless, for the processes studied in this paper four-fermion operators do not contribute, therefore the corresponding results in Refs.~\cite{Ferreira:2005dr,Ferreira:2006in} can be related to ours by simply using Eqs.~(\ref{ec:trans}).

Besides, for a better understanding of the implications of operator equalities we also write down the relations among four-fermion operator coefficients for FCN $tu$ interactions. The four-fermion operator Lagrangian in Ref.~\cite{Ferreira:2006xe} reads
\begin{eqnarray}
\mathcal{L}'_\text{4F} & = & \frac{g_s \gamma_{u_1}}{\Lambda^2}
\left[ 2 \, O_{uu}^{(8,3111)} + 2 \, O_{uu}^{(8,3122)} +
O_{ud}^{(8,3111)} + O_{ud}^{(8,3122)} + O_{ud}^{(8,3133)} \right] \notag \\
& & + \frac{g_s \gamma_{u_2}}{\Lambda^2}
\left[ O_{qu}^{(8,1131)} + O_{qu}^{(8,2132)} + O_{qu}^{(8,2231)} \right] \notag \\
& & + \frac{g_s \gamma_{u_3}}{\Lambda^2}
 \left[ - x O_{qq\epsilon}^{(8,3133)} - O_{qq\epsilon}^{(8,1311)}
  - O_{qq\epsilon}^{(8,1322)} - O_{qq\epsilon}^{(8,2312)} \right] + \text{h.c.} \,,
\end{eqnarray}
plus analogous terms for the $tc$ interactions, where
\begin{equation}
O_{qq\epsilon}^{(8,ijkl)} = (\bar q_{Li} \la u_{Rj}) ( [\bar q_{Lk} \epsilon]^T \la
d_{Rl} ) \,.
\end{equation}
Then, Eqs.~(\ref{ec:Oprel}) imply the relations
\begin{align}
& C_{uu}^{(8,3111)} = C_{uu}^{(8,3122)} = 2 g_s \gamma_{u_1} + i \frac{g_s}{2} (\atu-\aut^*) \,, \notag \\
& C_{ud}^{(8,3111)} = C_{ud}^{(8,3122)} = C_{ud}^{(8,3133)} = g_s \gamma_{u_1} + i \frac{g_s}{4} (\atu-\aut^*) \,, \notag \\
& C_{qu}^{(8,1131)} = C_{qu}^{(8,2132)} = g_s \gamma_{u_2} + i \frac{g_s}{6} (\atu-\aut^*) \,, \notag \\
& C_{qu}^{(8,2231)} = g_s \gamma_{u_2} \,, \notag \\ 
& C_{qu}^{(1,1131)} = C_{qu}^{(1,2132)} = -i \frac{8 g_s}{9} (\atu-\aut^*) \,, \notag \\
& C_{qq\epsilon}^{(8,3133)} = - x g_s \gamma_{u_3} \,, \notag \\
& C_{qq\epsilon}^{(8,1311)} = C_{qq\epsilon}^{(8,1322)} = C_{qq\epsilon}^{(8,2312)} = - g_s \gamma_{u_3} \,.
\end{align}
In view of these equations, some clarifications are in order:
\begin{itemize}
\item[(i)] The assumption that several operator coefficients in $\mathcal{L}'_\text{4F}$ are equal is broken for the case of $O_{qu}^{(8,1131)}$, $O_{qu}^{(8,2132)}$ and $O_{qu}^{(8,2231)}$ by the use of the equations of motion, {\em i.e.} the application of Eqs.~(\ref{ec:Oprel}). We must also note that actually, $\mathcal{L}'_\text{4F}$ does not contain 12 (independent) gauge-invariant operators but only three, given by the three linear combinations with arbitrary coefficients $\gamma_{u_1}$, $\gamma_{u_2}$, $\gamma_{u_3}$.
\item[(ii)] The operator selection in $\mathcal{L}'_\text{4F}$ (discarding other 22 operators which also contribute to the same processes) is not consistent with the use of the equations of motion: the operators  $O_{qu}^{(1,1131)}$ and $O_{qu}^{(1,2132)}$ are not present in
$\mathcal{L}'_\text{4F}$ but appear when Eqs.~(\ref{ec:Oprel}) are applied.
\end{itemize}
For these reasons, the $tj$ cross sections in Ref.~\cite{Ferreira:2006xe} involving four-fermion operator contributions are not invariant under the operator replacements in Eqs.~(\ref{ec:Oprel}), while the ones for $t$, $\gamma t$ and $Zt$ are. This, of course, is not due to misuse of the equations of motion but concerns the too restrictive operator selection in $\mathcal{L}'_\text{4F}$.

\section{Top FCN decays and direct top production}
\label{sec:3}

\begin{figure}[htb]
\begin{center}
\begin{tabular}{ccc}
\epsfig{file=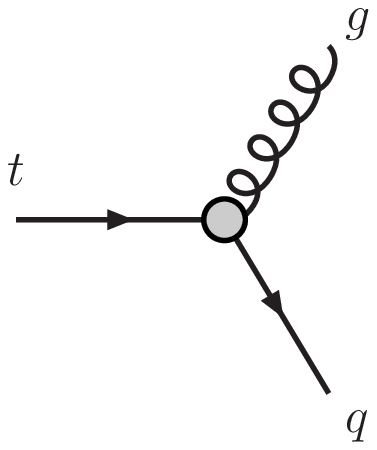,height=3cm,clip=} & \hspace{4cm} &
\epsfig{file=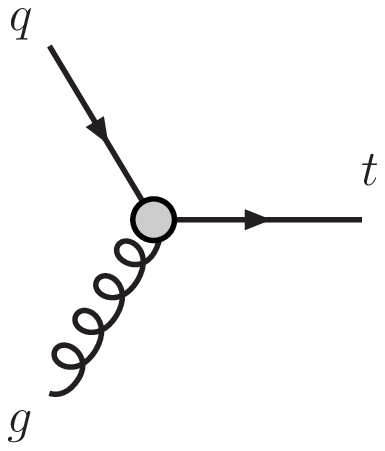,height=3cm,clip=} \\[2mm]
\end{tabular}
\caption{Left: Feynman diagram for top FCN decay $t \to qg$. Right: diagram for direct $t$ production.}
\label{fig:diag-t}
\end{center}
\end{figure}

In terms of the anomalous couplings defined by Eq.~(\ref{ec:gtu}),
the partial widths for top FCN decays $t \to qg$ (see Fig.~\ref{fig:diag-t}) are~\cite{AguilarSaavedra:2004wm}
\begin{equation}
\Gamma(t \to qg) = \frac{4}{3} \alpha_s m_t \left[ |\gql|^2 + |\gqr|^2 \right] \,.
\label{ec:tdec}
\end{equation}
We have neglected the up and charm quark masses. The proportionality
$\Gamma(t \to qg) \propto \left[ |\gql|^2 + |\gqr|^2 \right]$ of the partial widths is understood from simple arguments: (i) the amplitudes corresponding to $\gql$ and $\gqr$ do not interfere and a possible $\gql {\gqr}^*$ term is thus absent; (ii) the top quark polarisation is averaged and $q$, $g$ polarisations summed, so that the coefficients multiplying $|\gql|^2$ and $|\gqr|^2$ are equal.

Writing the anomalous couplings in terms of the effective operator coefficients of our minimal basis, and using Eqs.~(\ref{ec:trans}) to relate them to the one in Ref.~\cite{Ferreira:2005dr},
we can recover the expressions for the partial widths given in Eq.~(9) of that reference,
\begin{eqnarray}
\Gamma(t \to qg) & = & \frac{m_t^3}{12 \pi \Lambda^4} \left \{
m_t^2 |\atq+\aqt^*|^2 + 16 \hat v^2 \left[ |\btq|^2 + |\bqt|^2 \right] \right. \notag \\[1mm]
& & \left.  + 8 \hat v m_t \,\IM \btq (\atq^* + \aqt) \right \} \,,
\end{eqnarray}
with $\hat v=v/\sqrt 2 = 174$ GeV and $\Lambda$ in GeV. Hence, both expressions are equivalent, as implied by gauge invariance, but our Eq.~(\ref{ec:tdec}) is much simpler and contains the same physics.

The amplitudes for direct production $gq \to t$ are related to the ones for top FCN decay by crossing symmetry (see Fig.~\ref{fig:diag-t}), hence the cross sections are also proportional to the respective
factors $\left[ |\gql|^2 + |\gqr|^2 \right]$, without interference terms. This obviously implies that
\begin{align}
& \sigma(gu \to t) \propto \Gamma(t \to gu) \,, \notag \\
& \sigma(gc \to t) \propto \Gamma(t \to gc) \,,
\end{align}
and the same for antiquarks, as it was pointed out in Ref.~\cite{Ferreira:2005dr}. 
The cross sections for the different processes can be calculated with a numerical integration in each case. Using CTEQ6L1 parton distribution functions (PDFs) and setting $Q^2 = m_t^2$, we find for LHC with a CM energy of 14 TeV
\begin{align}
& \sigma(gu \to t) = 2.162 \times 10^6 \left[ |\gul|^2 + |\gur|^2 \right] ~\text{pb} \,, \notag \\
& \sigma(g \bar u \to \bar t) = 5.153 \times 10^5 \left[ |\gul|^2 + |\gur|^2 \right] ~\text{pb} \,, \notag \\
& \sigma(gc \to t) = 2.985 \times 10^5 \left[ |\gcl|^2 + |\gcr|^2 \right] ~\text{pb} \,,
\label{ec:xsec-t}
\end{align}
and $\sigma(g\bar c \to \bar t) = \sigma(gc \to t)$.
The top quark decay (with a left-handed coupling) introduces an asymmetry of 0.8\% at most between the numerical coefficients of left- and right-handed couplings. This difference can be ignored to a good approximation in the cross section calculations.
Writing the anomalous couplings in terms of effective operator coefficients and
using Eqs.~(\ref{ec:trans}) we can recover the cross section expressions in Ref.~\cite{Ferreira:2005dr}, for example in their Eqs.~(17) and (18),
\begin{eqnarray}
 \sigma(gu \to t) & = & \frac{1}{\Lambda^4} \left \{  342 |\atu+\aut^*|^2 + 5413 \left[ |\btu|^2
+ |\but|^2 \right] \right. \notag \\[1mm]
& & \left. + 2722 \, \IM \btu (\atu^* + \aut) \right \} ~\text{pb} \,, \notag \\
\sigma(gc \to t) & = & \frac{1}{\Lambda^4} \left \{  47.2 |\atc+\act^*|^2 + 747 \left[ |\btc|^2
+ |\bct|^2 \right] \right. \notag \\[1mm]
& & \left. + 376 \, \IM \btc (\atc^* + \act) \right \} ~\text{pb} \,,
\end{eqnarray}
with $\Lambda$ in TeV. Our numerical coefficients are 7\% larger for the former process and 7\% smaller for the latter due to the different PDFs used. Thus, apart from the precise numerical coefficients resulting from the integration, our cross sections in Eqs.~(\ref{ec:xsec-t}) are completely equivalent to those in Ref.~\cite{Ferreira:2005dr}. For LHC at 7 TeV the total cross sections are
\begin{align}
& \sigma(gu \to t) = 8.013 \times 10^5 \left[ |\gul|^2 + |\gur|^2 \right] ~\text{pb} \,, \notag \\
& \sigma(g \bar u \to \bar t) = 1.428 \times 10^5 \left[ |\gul|^2 + |\gur|^2 \right] ~\text{pb} \,, \notag \\
& \sigma(gc \to t) = 7.424 \times 10^4 \left[ |\gcl|^2 + |\gcr|^2 \right] ~\text{pb} \,,
\label{ec:xsec-t7}
\end{align}
and for Tevatron
\begin{align}
& \sigma(gu \to t) = 4.065 \times 10^4 \left[ |\gul|^2 + |\gur|^2 \right] ~\text{pb} \,, \notag \\
& \sigma(gc \to t) = 2247 \left[ |\gcl|^2 + |\gcr|^2 \right] ~\text{pb} \,,
\label{ec:xsec-t2}
\end{align}
with equal cross sections for $t$ and $\bar t$ production, as corresponds to a $p\bar p$ collider.

\section{$\gamma t$ production}
\label{sec:4}

In this section we first study $\gamma t$ production with the top quark treated as an on-shell final state particle, and then we present the results taking into account the top quark decay. In the latter case for brevity we restrict our study to leptonic final states $\gamma \ell \nu b$, which are the most interesting ones for the experimental observation of this process.

\subsection{$\gamma t$ production with $t$ on-shell} 

\begin{figure}[htb]
\begin{center}
\begin{tabular}{ccc}
\epsfig{file=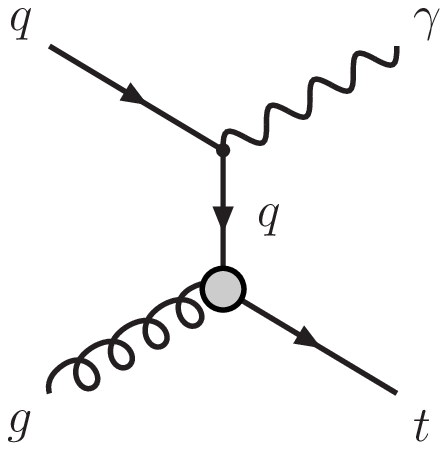,height=3cm,clip=} & \quad \quad \quad \quad \quad &
\epsfig{file=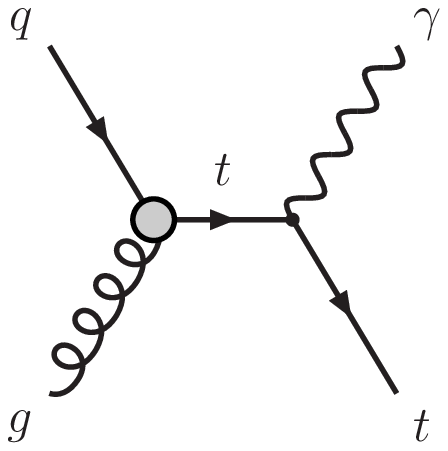,height=3cm,clip=}
\end{tabular}
\caption{Feynman diagrams contributing to $\gamma t$ production mediated by FCN $gtq$ couplings.}
\label{fig:diag-At4}
\end{center}
\end{figure}

The diagrams contributing to $\gamma t$ production mediated by $gtq$ couplings are depicted in Fig.~\ref{fig:diag-At4}.
They are related to the one for $gq \to t$ in Fig.~\ref{fig:diag-t} (right) by the insertion of a photon in one of the two quark legs. For this process the proportionality between cross sections and branching ratios, found in Ref.~\cite{Ferreira:2006in} by explicit calculation, is also evident from symmetry arguments: (i) since the photon does not change the fermion chirality, the amplitudes corresponding to the two operators do not interfere and a $\gql {\gqr}^*$ term is not present; (ii) the vector coupling of the photon conserves parity, thus the coefficient of both terms is the same,
$\sigma (gq \to \gamma t) \propto \left[ |\gql|^2 + |\gqr|^2 \right]$. Therefore, the relations
\begin{eqnarray}
\sigma(gu \to \gamma t) & \propto & \Gamma(t \to gu) \,, \notag \\
\sigma(gc \to \gamma t) & \propto & \Gamma(t \to gc) 
\end{eqnarray}
also hold, for quarks as well as for antiquarks.

\begin{figure}[t]
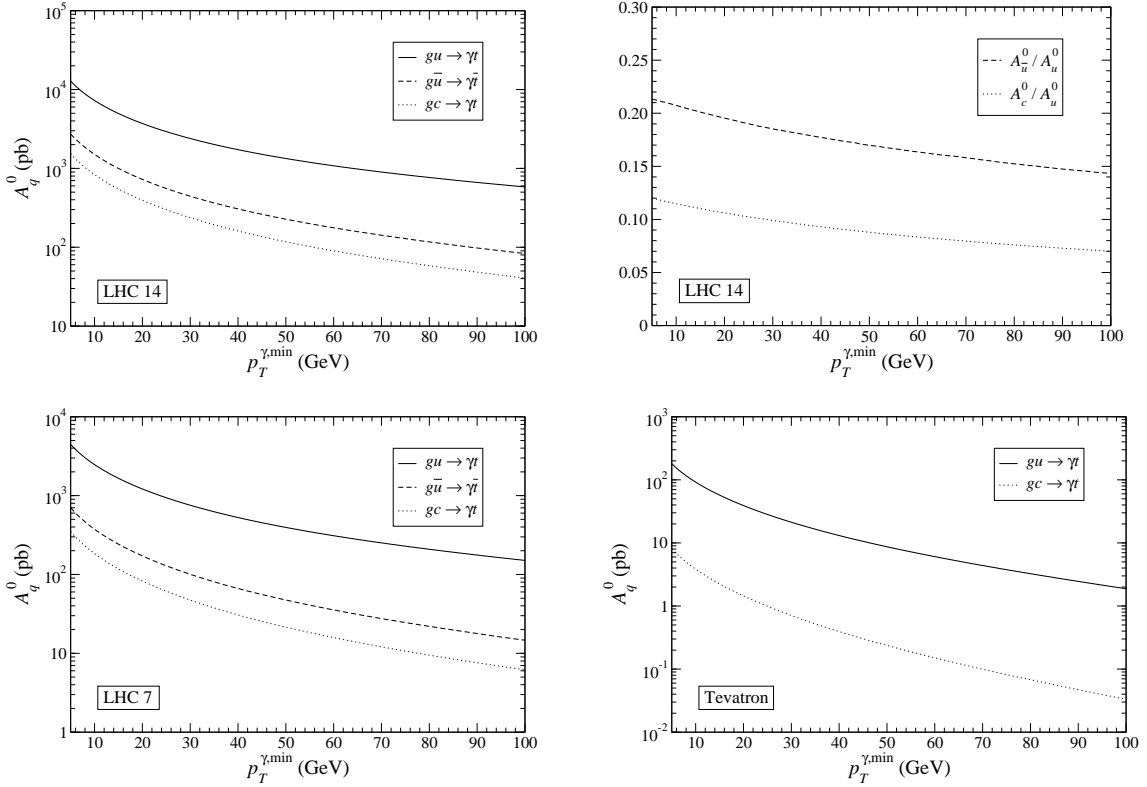

\begin{center}
\begin{tabular}{ccc}
\epsfig{file=Figs/cross-At4.eps,height=5cm,clip=} & \quad &
\epsfig{file=Figs/R-At4.eps,height=5cm,clip=} \\[2mm]
\epsfig{file=Figs/cross7-At4.eps,height=5cm,clip=} & \quad &
\epsfig{file=Figs/cross2-At4.eps,height=5cm,clip=}
\end{tabular}
\caption{Up, left: proportionality factors $A_u^0$, $A_{\bar u}^0$, $A_c^0$ for $\gamma t$ and $\gamma \bar t$ production ($2 \to 2$ process) for LHC at 14 TeV. Up, right: their ratios. Down: the same factors for LHC at 7 TeV (left) and for Tevatron (right).}
\label{fig:cross-At4}
\end{center}
\end{figure}

The process of $\gamma t$ production through $gtq$ couplings has a divergence from the $t$-channel diagram in Fig.~\ref{fig:diag-At4} when the photon is collinear to the initial quark. To avoid this divergence one can impose a minimum transverse momentum cut $p_T^{\gamma,\text{min}}$ for the emitted photons. We have
\begin{align}
& \sigma(gu \to \gamma t) = A_u^0 \left[ |\gul|^2 + |\gur|^2 \right] \,, \notag \\
& \sigma(g \bar u \to \gamma \bar t) = A_{ \bar u}^0 \left[ |\gul|^2 + |\gur|^2 \right] \,, \notag \\
& \sigma(gc \to \gamma t) = A_c^0 \left[ |\gcl|^2 + |\gcr|^2 \right] \,,
\label{ec:xsec-At}
\end{align}
with $\sigma(g \bar c \to \gamma \bar t) = \sigma(gc \to \gamma t)$.
The proportionality factors $A_u^0$, $A_{\bar u}^0$, $A_c^0$ are given in Fig.~\ref{fig:cross-At4} for LHC and Tevatron, as a function of the $p_T^{\gamma,\text{min}}$ cut. They have been evaluated using CTEQ6L1 PDFs with $Q^2 = m_t^2$. 
On the up, right panel we show the ratios $A_{\bar u}^0 / A_u^0$ and $A_c^0 / A_u^0$ for LHC at 14 TeV, used to cross-check our calculations. 

\newpage
In order to compare our results with previous work we set $p_T^{\gamma,\text{min}} = 15$ GeV, obtaining for LHC at 14 TeV
\begin{eqnarray}
\sigma_{p_T^\gamma > 15}(gu \to \gamma t) & = & 4966 \left[ |\gul|^2 + |\gur|^2 \right] ~\text{pb} \notag \\
& = & 180.6 \, \Gamma(t \to ug) ~\text{pb/GeV} \,, \notag \\
\sigma_{p_T^\gamma > 15}(g \bar u \to \gamma \bar t) & = & 998.6 \left[ |\gul|^2 + |\gur|^2 \right] ~\text{pb} \notag \\
& = & 36.3 \, \Gamma(t \to ug) ~\text{pb/GeV} \,, \notag \\
\sigma_{p_T^\gamma > 15}(gc \to \gamma t) & = & 547.4 \left[ |\gcl|^2 + |\gcr|^2 \right] ~\text{pb} \notag \\
& = & 19.9 \, \Gamma(t \to ug) ~\text{pb/GeV} \,,
\label{ec:xsec15-At4}
\end{eqnarray}
corresponding to Eqs.~(20) of Ref.~\cite{Ferreira:2006in}. 
For the second process we find a good numerical agreement (our value is around 10\% larger) but not for the first one. The discrepancy seems to be due to a typo in Ref.~\cite{Ferreira:2006in}, because for $p_T^{\gamma,\text{min}} = 15$ GeV our ratio $A_{\bar u}^0 / A_u^0 = 0.2$ (see Fig.~\ref{fig:cross-At4}) is not far from the ratio of direct $t$ production cross sections, as expected, while Ref.~\cite{Ferreira:2006in} finds a much smaller value of 0.14.\footnote{If the proportionality $\sigma_{p_T^\gamma > 15}(gu \to \gamma t) = 228 \, \Gamma(t \to ug)~\text{pb/GeV}$ in Eq.~(20) of Ref.~\cite{Ferreira:2006in} is understood as $\sigma_{p_T^\gamma > 15}(gu \to \gamma t) = 228 \, \mathrm{Br}(t \to ug)~\text{pb}$ and we use their value for the total top width $\Gamma_t = 1.42$ GeV, the resulting proportionality $\sigma_{p_T^\gamma > 15}(gu \to \gamma t) = 160 \, \Gamma(t \to ug)~\text{pb/GeV}$ is in good agreement with our calculation, and the ratio of cross sections $A_{\bar u}^0 / A_u^0 = 0.2$ perfectly coincides with our value.}

\subsection{$\gamma \ell \nu b$ production}

\begin{figure}[htb]
\begin{center}
\begin{tabular}{ccccc}
\epsfig{file=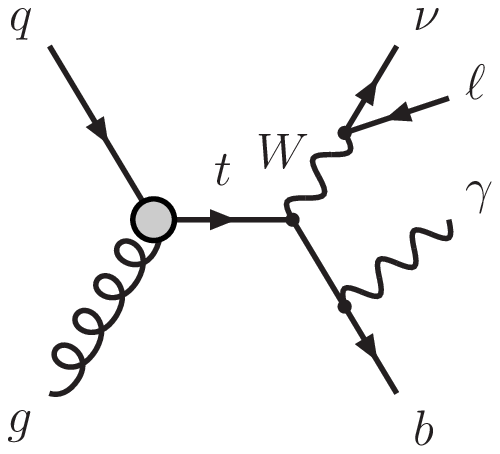,height=3cm,clip=} & \quad \quad \quad \quad &
\epsfig{file=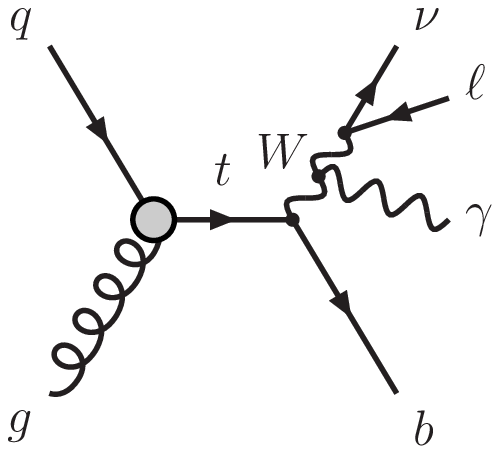,height=3cm,clip=} & \quad \quad \quad \quad &
\epsfig{file=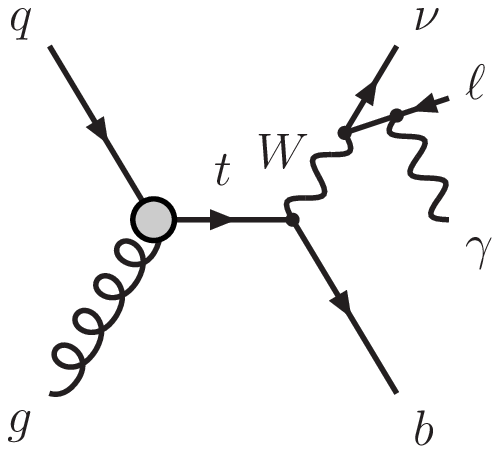,height=3cm,clip=}
\end{tabular}
\caption{Additional Feynman diagrams contributing to $\gamma \ell \nu b$ production mediated by FCN $gtq$ couplings.}
\label{fig:diag-Atextra}
\end{center}
\end{figure}

\begin{figure}[htb]
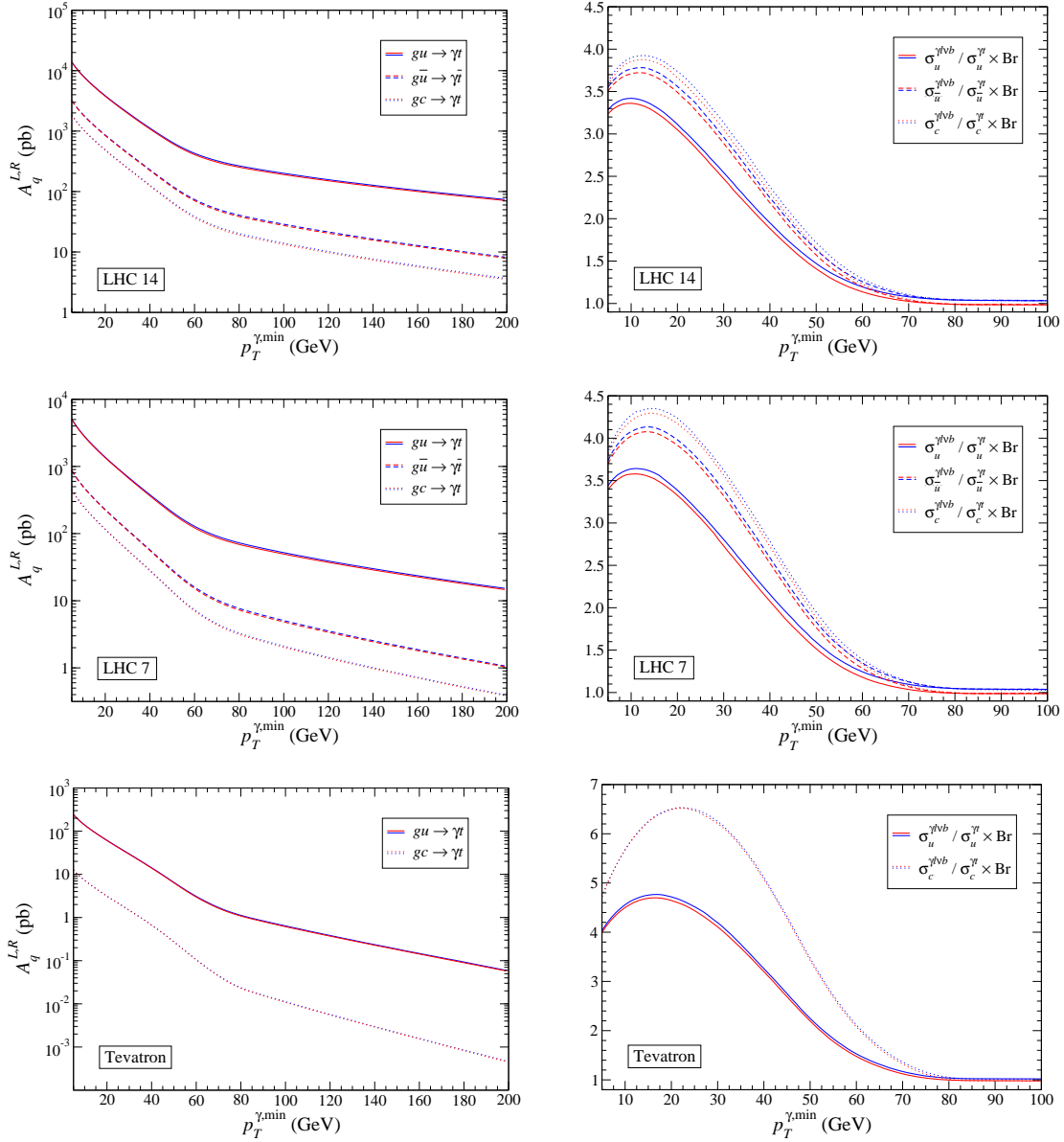

\begin{center}
\begin{tabular}{ccc}
\epsfig{file=Figs/cross-At.eps,height=5cm,clip=} & \quad &
\epsfig{file=Figs/R6to4-At.eps,height=5cm,clip=} \\[2mm]
\epsfig{file=Figs/cross7-At.eps,height=5cm,clip=} & \quad &
\epsfig{file=Figs/R6to4-7-At.eps,height=5cm,clip=} \\[2mm]
\epsfig{file=Figs/cross2-At.eps,height=5cm,clip=} & \quad &
\epsfig{file=Figs/R6to4-2-At.eps,height=5cm,clip=}
\end{tabular}
\caption{Left: proportionality factors $A_u^{L,R}$, $A_{\bar u}^{L,R}$, $A_c^{L,R}$ defined in Eqs.~(\ref{ec:xsec-At6}) for $\gamma \ell \nu b$ production. Right: ratios of cross sections including all diagrams or only the resonant $\gamma t$ production. The blue (upper) and red (lower) lines correspond to left- and right-handed couplings, respectively.}
\label{fig:Atcross}
\end{center}
\end{figure}

When the top decay $t \to W b \to f \bar f' b$ is included the diagrams in Fig.~\ref{fig:diag-At4} are not the only ones contributing to the $\gamma f \bar f' b$ final state, but additional ones appear where the photon is radiated from the top decay products. For brevity, we will restrict ourselves to the leptonic final state $\gamma \ell \nu b$, which has much smaller backgrounds and is the most interesting from the experimental point of view. The additional diagrams for $g q \to \gamma \ell \nu b$ are depicted in Fig.~\ref{fig:diag-Atextra}.
They correspond to direct top production and photon radiation off the $b$ quark, $W$ boson and charged lepton, respectively. The importance of these diagrams is evident if one considers the large direct top production cross section in Eqs.~(\ref{ec:xsec-t}) and the branching ratio
$\text{Br}(t \to \gamma W b) \sim 3.5 \times 10^{-3}$~\cite{Mahlon:1998fr,Mele:1999zx}.
The production cross sections, summing over $\ell = e,\mu,\tau$, can be written as 
\begin{align}
& \sigma(gu \to \gamma \ell \nu b) = A_u^L |\gul|^2 + A_u^R |\gur|^2 \,, \notag \\
& \sigma(g \bar u \to \gamma \ell \nu \bar b) = A_{ \bar u}^L |\gul|^2 + A_{ \bar u}^R |\gur|^2 \,, \notag \\
& \sigma(gc \to \gamma \ell \nu b) = A_c^L |\gcl|^2 + A_c^R |\gcr|^2 \,,
\label{ec:xsec-At6}
\end{align}
where now the coefficients of left- and right-handed couplings are not equal, since parity is violated in the top quark decay vertex (as well as in the $W$ decay to $\ell \nu$). 
The numerical values of these coefficients, as a function of the minimum photon transverse momentum cut $p_T^{\gamma,\text{min}}$, are presented in Fig.~\ref{fig:Atcross} (left) for LHC and Tevatron.
In order to remove the divergence from the photon emission off the charged lepton, we have required a lego-plot separation $\Delta R_{\gamma \ell} > 0.4$. The importance of the contributions from the diagrams in Fig.~\ref{fig:diag-Atextra} to the cross section can be appreciated by calculating the ratio
\begin{equation}
\frac{\sigma(\gamma \ell \nu b)}{\sigma(\gamma t) \times \text{Br}(t \to \ell \nu b)} \,,
\end{equation}
where $\text{Br}(t \to \ell \nu b) \simeq 1/3$.  This ratio is plotted in Fig.~\ref{fig:Atcross} (right) for the different processes. We observe that the ``new'' contributions where the photon is radiated from the top quark decay products dominate even up to a relatively large cut $p_T^{\gamma,\text{min}} \lesssim 40$. Note also the quick decrease of the cross sections with $p_T^{\gamma,\text{min}}$, up to $p_T^{\gamma,\text{min}} \simeq 70~\text{GeV}$ (compare with Fig.~\ref{fig:cross-At4}). This is due to the suppression of the amplitudes in Fig.~\ref{fig:diag-Atextra} for increasing photon transverse momentum. The relative size of the new contributions depends on the minimum photon $p_T$ as well as on the CM energy and initial state. For $p_T^{\gamma,\text{min}} = 20$ GeV the enhancement is by factors $3-4.5$ for LHC and $4.5 - 6.5$ at Tevatron.

\begin{figure}[t]
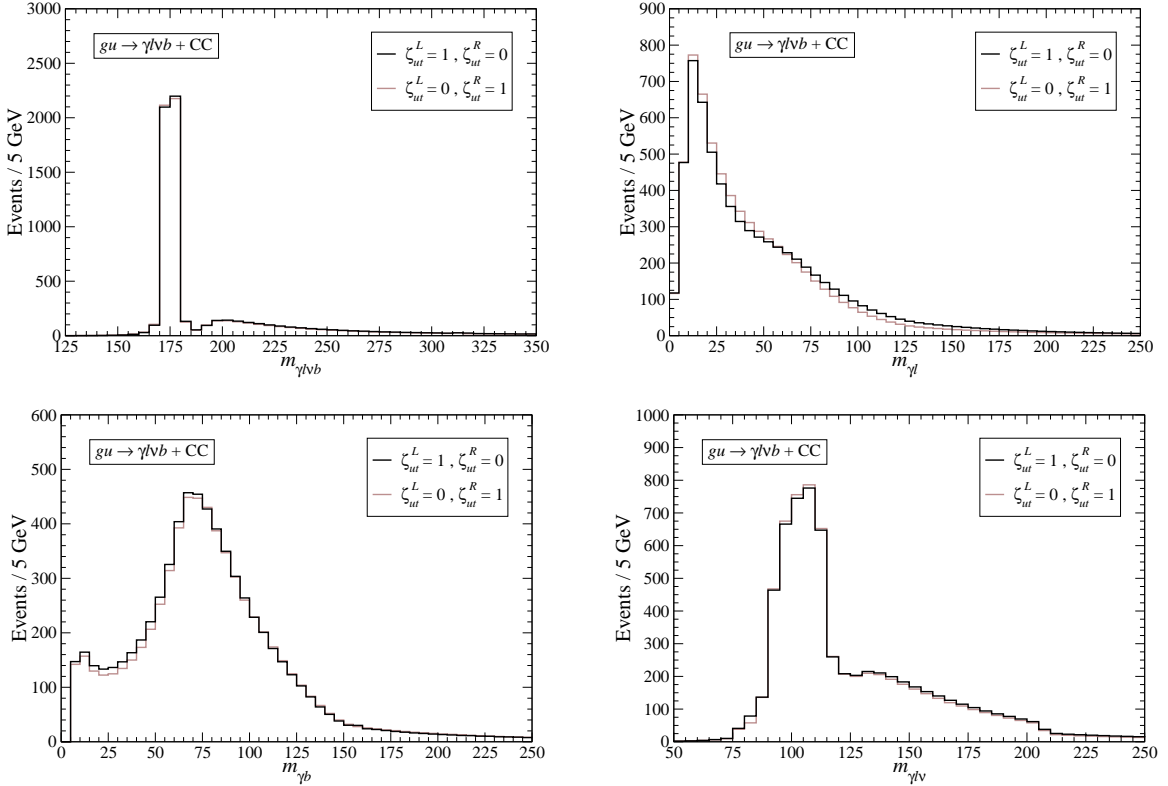

\begin{center}
\begin{tabular}{ccc}
\epsfig{file=Figs/mtot-At.eps,height=5cm,clip=} & \quad &
\epsfig{file=Figs/mlA-At.eps,height=5cm,clip=} \\[2mm]
\epsfig{file=Figs/mbA-At.eps,height=5cm,clip=} & \quad &
\epsfig{file=Figs/mWA-At.eps,height=5cm,clip=}
\end{tabular}
\caption{Invariant mass distributions for LHC at 14 TeV, with $p_T^{\gamma,\text{min}} = 15$ GeV: $\gamma \ell \nu b$ (up, left) $\gamma \ell$ (up, right); $\gamma b$ (down, left); $\gamma \ell \nu$ (down, right).}
\label{fig:Atdist}
\end{center}
\end{figure}

We follow our comparison with previous work by setting $p_T^{\gamma,\text{min}} = 15$ GeV, to obtain
\begin{eqnarray}
\sigma_{p_T^\gamma > 15}(gu \to \gamma \ell \nu b) & = & 5545 |\gul|^2 + 5431 |\gur|^2 ~\text{pb} \notag \\
& \simeq & 3.3 \, \sigma_{p_T^\gamma > 15}(gu \to \gamma t) \times \text{Br} (t \to \ell \nu b) \,, \notag \\
\sigma_{p_T^\gamma > 15}(g \bar u \to \gamma \ell \nu \bar b) & = & 1250 |\gul|^2 + 1229 |\gur|^2 ~\text{pb} \notag \\
& \simeq & 3.7 \, \sigma_{p_T^\gamma > 15}(g \bar u \to \gamma \bar t) \times \text{Br} (t \to \ell \nu b) \,, \notag \\
\sigma_{p_T^\gamma > 15}(gc \to \gamma \ell \nu b) & = & 714.8 |\gcl|^2 + 706.7 |\gcr|^2 ~\text{pb} \notag \\
& \simeq & 3.9 \, \sigma_{p_T^\gamma > 15}(gc \to \gamma t) \times \text{Br} (t \to \ell \nu b) \,,
\label{ec:xsec15-At}
\end{eqnarray}
for LHC at 14 TeV.
Thus, we see that approximating the process by on-shell $\gamma t$ production with subsequent top decay~\cite{Ferreira:2006in} only accounts for a $25\%-30\%$ of the cross section. This is clearly seen in Fig.~\ref{fig:Atdist} (up, left), where we plot the total $\gamma \ell \nu b$ invariant mass distribution. The sharp peak corresponds to direct top production with $t \to \gamma \ell \nu b$, while the right part with $m_{\gamma \ell \nu b} \gtrsim 190$ GeV is on-shell $\gamma t$ production with $t \to \ell \nu b$. It is also interesting to examine the invariant mass distributions of the photon and the charged lepton, the $b$ quark and the $W$ boson. These distributions show that the photon is mainly emitted from the former, due to its smaller mass, and that the divergence is well cut by the $\Delta R$ requirement.

Finally, it is important to remark that the contributions from direct top production and $t \to \gamma \ell \nu b$ are observable even when transverse momentum and isolation cuts are imposed. In order to show this, we have calculated the cross sections for LHC at 14 TeV with cuts
\begin{equation}
p_T^{\gamma,\ell,b} > 15~\text{GeV} \;, \quad \quad |\eta_{\gamma,\ell,b}| < 2.5
\;, \quad \quad \Delta R_{\gamma \ell,\gamma b,\ell b} > 0.4 \,,
\end{equation}
obtaining
\begin{eqnarray}
\sigma_\text{cut}(gu \to \gamma \ell \nu b) & \simeq & 2.6 \, \sigma_\text{cut}(gu \to \gamma t \to \gamma \ell \nu b) \,, \notag \\
\sigma_\text{cut}(g \bar u \to \gamma \ell \nu \bar b) & \simeq & 2.8 \, \sigma_\text{cut}(g \bar u \to \gamma \bar t \to \gamma \ell \nu b) \,, \notag \\
\sigma_\text{cut}(gc \to \gamma \ell \nu b) & \simeq & 2.9 \, \sigma_\text{cut}(gc \to \gamma t \to \gamma \ell \nu b) \,.
\label{ec:xsec-cut-At}
\end{eqnarray}

\newpage
\section{$Zt$ production}
\label{sec:5}

For better illustration, we first present in this section the results considering the $Z$ boson and top quark as on-shell particles, followed by the calculations including both decays.

\subsection{$Zt$ production with $Z$, $t$ on-shell} 

\begin{figure}[htb]
\begin{center}
\begin{tabular}{ccc}
\epsfig{file=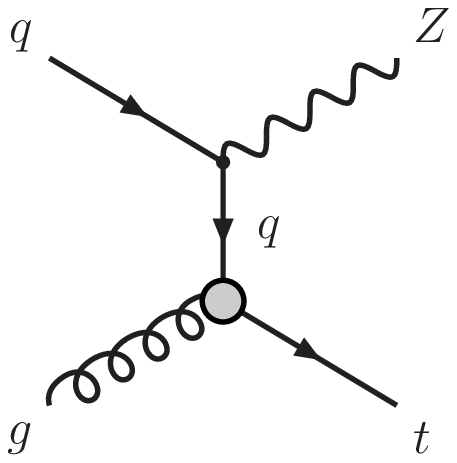,height=3cm,clip=} & \quad \quad \quad \quad \quad &
\epsfig{file=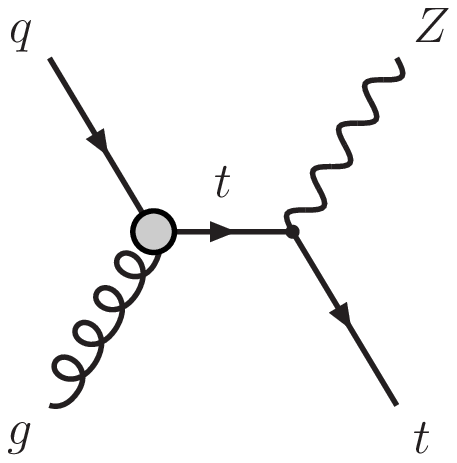,height=3cm,clip=}
\end{tabular}
\caption{Feynman diagrams contributing to $Zt$ production mediated by FCN $gtq$ couplings.}
\label{fig:diag-Zt4}
\end{center}
\end{figure}

This process is related to direct $t$ production by the insertion of a $Z$ boson in a quark line, see Fig.~\ref{fig:diag-Zt4}.
Because the $Z$ coupling conserves chirality, the amplitudes corresponding to the two operators do not interfere and the matrix element does not contain a 
$\gql {\gqr}^*$ term. However, in contrast to the previous cases the parity-violating $Z$ coupling 
to the quarks introduces an asymmetry between the coefficients of the two operators corresponding to different quark chiralities, so that the cross sections for $Zt$, $Z \bar t$ production are not proportional to branching ratios.\footnote{In Ref.~\cite{Ferreira:2006in} the absence of this proportionality is attributed to the fact that the $Z$ boson is massive. Clearly, the $Z$ boson mass does not have any relation with parity violation, and we have explicitly checked that with a massive $Z$ boson with a vector coupling the proportionality $\sigma (gq \to Z t) \propto \left[ |\gql|^2 + |\gqr|^2 \right]$ is restored.}

The cross sections are calculated using CTEQ6L1 PDFs with $Q^2 = m_t^2+M_Z^2$. We obtain for LHC at 14 TeV
\begin{align}
& \sigma(gu \to Z t) = 2.321 \times 10^4 |\gul|^2 + 2.378 \times 10^4 |\gur|^2 ~\text{pb} \,, \notag \\
& \sigma(g \bar u \to Z \bar t) = 2344 |\gul|^2 + 2445 |\gur|^2 ~\text{pb} \,, \notag \\
& \sigma(gc \to Z t) = 1060 |\gcl|^2 + 1111 |\gcr|^2 ~\text{pb} \,.
\label{ec:xsec-Zt}
\end{align}
We note that in this case it is not necessary to set a minimum $p_T$ cut because the cross sections are finite.  In order to compare with previous numerical results we 
use Eqs.~(\ref{ec:ztoC}) and (\ref{ec:trans}), obtaining
\begin{eqnarray}
 \sigma(gu \to Zt) & = & \frac{1}{\Lambda^4} \left \{  3.88 |\atu+\aut^*|^2 + 61.3 |\btu|^2
+ 62.8 |\but|^2 \right. \notag \\[1mm]
& & \left. + 30.8 \, \IM \btu (\atu^* + \aut) \right \} ~\text{pb} \,, \notag \\
 \sigma(g \bar u \to Z \bar t) & = & \frac{1}{\Lambda^4} \left \{  0.392 |\atu+\aut^*|^2 + 6.19 |\btu|^2 + 6.46 |\but|^2 \right. \notag \\[1mm]
& & \left. + 3.11 \, \IM \btu (\atu^* + \aut) \right \} ~\text{pb} \,, \notag \\
\sigma(gc \to Zt) & = & \frac{1}{\Lambda^4} \left \{  0.177 |\atc+\act^*|^2 + 2.80 |\btc|^2
+ 2.93 |\bct|^2 \right. \notag \\[1mm]
& & \left. + 1.41 \, \IM \btc (\atc^* + \act) \right \} ~\text{pb} \,,
\end{eqnarray}
with $\Lambda$ in TeV. These results agree well with Eqs.~(21) of Ref.~\cite{Ferreira:2006in}, being our cross sections 4\%, 8\% and 15\% smaller due to the different PDFs used. For LHC at 7 TeV we find
\begin{align}
& \sigma(gu \to Z t) = 3953 |\gul|^2 + 4121 |\gur|^2 ~\text{pb} \,, \notag \\
& \sigma(g \bar u \to Z \bar t) = 307.4 |\gul|^2 + 327.3 |\gur|^2 ~\text{pb} \,, \notag \\
& \sigma(gc \to Z t) = 124.1 |\gcl|^2 + 133.2 |\gcr|^2 ~\text{pb} \,,
\label{ec:xsec7-Zt}
\end{align}
and for Tevatron
\begin{align}
& \sigma(gu \to Z t) = 29.38 |\gul|^2 + 32.62 |\gur|^2 ~\text{pb} \,, \notag \\
& \sigma(g c \to Z t) = 539.5 |\gcl|^2 + 617.4 |\gcr|^2 ~\text{fb} \,.
\label{ec:xsec2-Zt}
\end{align}

\subsection{$\ell^{\prime+} \ell^{\prime-} \ell \nu b$ production}

We restrict ourselves to final states with leptonic decays, $W \to \ell \nu$ and $Z \to \ell^{\prime+} \ell^{\prime-}$, summing over all charged leptons. In our computations we also include the diagrams in which the $Z$ boson is emitted from one of the top quark decay products, finding that their contribution to the cross section is at the per mille level. The main effect of the top decay is to reduce slightly the relative difference between cross sections corresponding to left-handed ($\gul$, $\gcl$) and right-handed ($\gur$, $\gcr$) couplings. For LHC at 14 TeV we have
\begin{align}
& \sigma(gu \to \ell^{\prime+} \ell^{\prime-} \ell \nu b) = 785.9 |\gul|^2 + 795.3 |\gur|^2 ~\text{pb} \,, \notag \\
& \sigma(g \bar u \to \ell^{\prime+} \ell^{\prime-} \ell \nu \bar b) = 79.18 |\gul|^2 + 81.60 |\gur|^2 ~\text{pb} \,, \notag \\
& \sigma(gc \to \ell^{\prime+} \ell^{\prime-} \ell \nu b) = 35.80 |\gcl|^2 + 37.12 |\gcr|^2 ~\text{pb} \,.
\label{ec:xsec-Zt7}
\end{align}
For 7 TeV the cross sections are
\begin{align}
& \sigma(gu \to \ell^{\prime+} \ell^{\prime-} \ell \nu b) = 133.5 |\gul|^2 + 137.6 |\gur|^2 ~\text{pb} \,, \notag \\
& \sigma(g \bar u \to \ell^{\prime+} \ell^{\prime-} \ell \nu \bar b) = 10.37 |\gul|^2 + 10.96 |\gur|^2 ~\text{pb} \,, \notag \\
& \sigma(gc \to \ell^{\prime+} \ell^{\prime-} \ell \nu b) = 4.193 |\gcl|^2 + 4.443 |\gcr|^2 ~\text{pb} \,,
\label{ec:xsec7-Zt7}
\end{align}
and for Tevatron
\begin{align}
& \sigma(gu \to \ell^{\prime+} \ell^{\prime-} \ell \nu b) = 0.9868 |\gul|^2 + 1.091 |\gur|^2 ~\text{pb} \,, \notag \\
& \sigma(g c \to \ell^{\prime+} \ell^{\prime-} \ell \nu b) = 18.17 |\gcl|^2 + 20.83 |\gcr|^2 ~\text{fb} \,.
\label{ec:xsec2-Zt7}
\end{align}

\section{Summary}
\label{sec:6}

In the forthcoming years, top quark interactions will be explored with an unprecedented precision, searching for deviations with respect to SM predictions which may signal new physics. From a theoretical point of view, it is of great importance to describe these interactions using a minimal (but complete) set of independent parameters, in order to translate experimental data into useful constraints on the effective Lagrangian.

In this paper we have studied several processes of single top production at hadron colliders, mediated by strong FCN top interactions: direct $t$, $\gamma t$ and $Zt$. Within an effective operator formalism, all these processes can be described in full generality by only two $gtu$ and two $gtc$ anomalous couplings~\cite{AguilarSaavedra:2008zc}. We have explicitly calculated cross sections in terms of this minimal set of parameters, obtaining a new insight on the relations between the different cross sections and the FCN top decay width $t \to qg$. If top FCN processes are eventually detected, these relations will allow to test whether the origin of the FCN interactions is in the strong sector or not.

In our calculations we have included the top quark and $Z$ boson decay, implementing these processes in the generator {\tt Protos}~\cite{AguilarSaavedra:2008gt}. 
The most conspicuous difference between our complete calculations and previous ones has been found in $\gamma t$ production. We have shown that the largest contribution to the $\gamma \ell \nu b$ signal (and analogously, to hadronic final states) results from
direct $gq \to t$ production followed by $t \to \gamma \ell \nu b$ decay. This contribution dominates up to relatively large photon momenta, would be experimentally detectable and has not been taken into account in previous literature. 
For LHC at 14 TeV (7 TeV), the increase in cross sections amounts to a factor of three (four), and for Tevatron it is even larger, around $4.5-6.5$. Of course, extra photons can also be produced by radiation in SM background processes involving or not the production of top quarks. In principle, this is taken into account in present background calculations and, in any case, the size of these backgrounds will be directly determined from data. On the other hand, for $\gamma \ell \nu b$ production we have shown that an adequate signal modelling must include the photon radiation from top quark decay products. Moreover,
in order to take advantage of this ``new'' signal contribution in future searches, it will necessary to look not only to final states of a top quark plus a photon~\cite{delAguila:1999ec}, but also to $\gamma \ell \nu b$ signals reconstructing a top quark.

\section*{Acknowledgements}

I thank M. P\'erez-Victoria and R. Pittau for useful comments. This work has been partially supported by a MICINN Ram\'on y Cajal contract and by projects FPA2006-05294 (MICINN), FQM 101 and FQM 437 (Junta de Andaluc\'{\i}a),
CERN/FP/83588/2008 (FCT),
and by the European Community's Marie-Curie Research Training
Network under contract MRTN-CT-2006-035505 ``Tools and Precision
Calculations for Physics Discoveries at Colliders''.

\end{document}